\documentclass[twocolumn,pre,epsf,superscriptaddress,showpacs,amsmath,amssymb,floatfix]{revtex4}
\usepackage{amsfonts}
\usepackage{amssymb}
\usepackage{amsmath}
\usepackage{graphicx}
\usepackage{hyperref}
\usepackage{bm}
\usepackage{color}

\begin{document}
\title{Self propelled particle transport in regular arrays of rigid asymmetric obstacles}
\author{Fabricio Q. Potiguar}
\email{fqpotiguar@ufpa.br}
\affiliation{Universidade Federal do Par\'a, Faculdade de F\'\i sica, ICEN, Av. Augusto Correa, 1, Guam\'a, 66075-110, Bel\'em, Par\'a, Brazil}
\author{G. A. Farias}
\affiliation{Departamento de F\'isica, Universidade Federal do
Cear\'a, Caixa Postal 6030, Campus do Pici, 60455-760 Fortaleza,
Cear\'a, Brazil}
\author{W. P. Ferreira}
\email{wandemberg@fisica.ufc.br}
\affiliation{Departamento de F\'isica, Universidade Federal do
Cear\'a, Caixa Postal 6030, Campus do Pici, 60455-760 Fortaleza,
Cear\'a, Brazil}

\begin{abstract}
We report numerical results which show the achievement of net transport of self-propelled particles (SPP) in the presence of a two-dimensional regular array of convex, either symmetric or asymmetric, rigid obstacles. The repulsive inter-particle (soft disks) and particle-obstacle interactions present no alignment rule. We find that SPP present a vortex-type motion around convex symmetric obstacles even in the absence of hydrodynamic effects. Such a motion is not observed for a single SPP, but is a consequence of the collective motion of SPP around the obstacles. An steady particle current is spontaneously established in an array of non-symmetric convex obstacle (which presents no cavity in which particles may be trapped in), and in the absence of an external field. Our results are mainly a consequence of the tendency of the self-propelled particles to attach to solid surfaces.
\end{abstract}


\maketitle
\section{Introduction}
Self-Propelled particles (SPP), also called swimmers, are entities that consume internal energy to generate motion \cite{toner05,ramaswamy10,vicsek12,marchetti12}. They are usually associated with motile microorganisms, artificial (Janus) micro-particles and flocking animals. Physical models that simulate such particles are divided between flocking (Vicsek model) \cite{vicsek95,drocco12-02}, and angular Brownian motion (ABM) types \cite{fily12} (which also includes run-and-tumble dynamics - RTD \cite{wan08,tailleur09}). Among the characteristics of these systems are the spontaneous appearance of motion orientational order \cite{vicsek95,chate06} and giant number fluctuations \cite{fily12,chate06,ramaswamy03}. It was seen that SPP are capable of turning gears and produce net work on large objects \cite{angelani09,dileonardo10,sokolov10,li13}, provided that there is an intrinsic asymmetry in such objects. Also, it is possible to separate SPP based on their rotational diffusion \cite{mijalkov13}. In addition to these phenomena, particle motion rectification was shown to occur when SPP are in the presence of funnel-shaped channels \cite{galajda07,wan08,drocco12-02,ghosh13,reichhardt13}. It was also shown that self-propelled rods can be trapped by moving barriers similar to funnel channels \cite{kaiser13}. Finally, Volpe {\em et al}. \cite{volpe11} showed that it is possible to sort swimmers using a periodic array of convex obstacles (ellipses) and an external drift force. 

In all these investigations medium asymmetry is a crucial ingredient for rectification to take place, and it is due to the broken time-symmetry \cite{cates12} in particle-obstacle interactions. In this Letter, we report a rectification effect similar to the one seen in Refs. [\onlinecite{galajda07,wan08,drocco12-02,ghosh13,reichhardt13}] for concave obstacles, but here employing periodic arrays of either symmetric (circles) or asymmetric (half-circles) {\it convex} obstacles (which presents no cavity in which particles may be trapped in), {\em without} the influence of any external drift force. We argue that the use of convex obstacles produces two distinct types of steady states: $i$) in the half-circle case, there is transport of particles with a non-zero mean drift velocity (also found in arrays of funnel objects \cite{reichhardt13}) and a constant density profile; $ii$) in the circle case, we observe a variable particle density profile and a zero drift velocity (no transport). Here, we will show only detailed results regarding the half-circle case, postponing the circle case for future work. Our results point to the possibility to devise sorting devices based on regular arrays of solid, convex obstacles.

\section{Model}
 Our model is of ABM type and consists of a two-dimensional (2D) system with $N$ swimmers in a $L \times L$ box, in which there is an array of $N_0$ static obstacles arranged in a square lattice with unit cell length (UCL) $a$. The obstacles are circles or half-circles of diameter $D$. Unless it is explicitly stated the normal direction to the flat side of the half-circles is the $+x$ direction. The swimmers are modeled as soft disks of diameter $d$, which interact through linear springs of stiffness $\kappa$. There is no specific inter-particle and particle-obstacle alignment rules \cite{baskaran08}. The swimmers move with a self-propelling velocity ${\bf v}_i=v_0\cos\theta_i(t){\bf i}+v_0\sin\theta_i(t){\bf j}$, whose random direction, $\theta_i(t)$, is proportional to a Gaussian white noise $\eta_i(t)$, which satisfies $\left<\eta_i(t)\right>=0$ and $\left<\eta_i(t)\eta_j(t^\prime)\right>=(2\eta\Delta t)^{1/2}\delta_{ij}\delta(t-t^\prime)$, with $\eta$ the noise intensity, and $\Delta t$ the time step. The particles follow a dynamics similar to the one presented in Ref. [\onlinecite{fily12}], i. e., there is no thermal Brownian motion. Interactions with obstacles are also of the linear spring form, but with a stiffness constant $\kappa_0>>\kappa$, in order to approximate the rigid body limit. The equations of motion for swimmer $i$ are written, in the overdampped case, as:
\begin{equation}
  \label{eq_1}
  \frac{\partial{\bf r}_i}{\partial t}={\bf v}_i+\mu{\bf F}_i,~~~~~\frac{\partial\theta_i}{\partial t}=\eta_i(t),
\end{equation}
\noindent where $\mu$ is the particle motility, ${\bf F}_i=\sum_{j}{\bf F}_{ij}$ is the total force in particle $i$ (sum is over $j$ particles and obstacles), ${\bf F}_{ij}=\kappa\alpha_{ij}{\bf \hat{r}}_{ij}$, if $\alpha_{ij}>0$ (${\bf F}_{ij}=0$ otherwise), and $\alpha_{ij}=\frac{1}{2}(d_i+d_j)-r_{ij}$ is the overlap distance between disk $i$ and object (disk or obstacle) $j$, $d_i=d$, and $d_j=d$ ($d_j=D$) for particle-particle (particle-obstacle) contact (for a contact with the flat side of a half-circle, $d_j=0$), and $r_{ij}$ is the distance between $i$ and $j$. Lengths are given in terms of the particle diameter $d$ and the time unit is set by $v_0=1$. Other parameter values are $L=100$, $\kappa=10$, $\kappa_0=1000$, $\mu=1$, $\Delta t=0.001$.  In all simulations we employed periodic boundary conditions (PBC) in both $x-$ and $y-$directions. The equations of motion are integrated using a second order, stochastic Runge-Kutta algorithm \cite{honeycutt92}. 
The rectification effect is characterized by the mean drift velocity $\left<v_i\right>$ ($i=x,y$), which is studied in terms of the UCL of the obstacle lattice ($a$), the obstacle size ($D$), the angular noise magnitude ($\eta$), and the area fraction ($\phi$). The latter is the ratio between the area occupied by the SPP and the area available to them, i. e., $\phi = N\pi d^2/[4(L^2-S_T)]$, where $S_T$ is the area covered by the obstacles. Notice that $\phi$ is related to the density $n$ of SPP, i.e. $\phi=\pi d^{2} n$.

\section{Results and discussions}
%
\subsection{Arrays of symmetric and asymmetric obstacles}
\begin{figure}[h]
  \centering
  \begin{tabular}{cc}
    \includegraphics[scale=0.2]{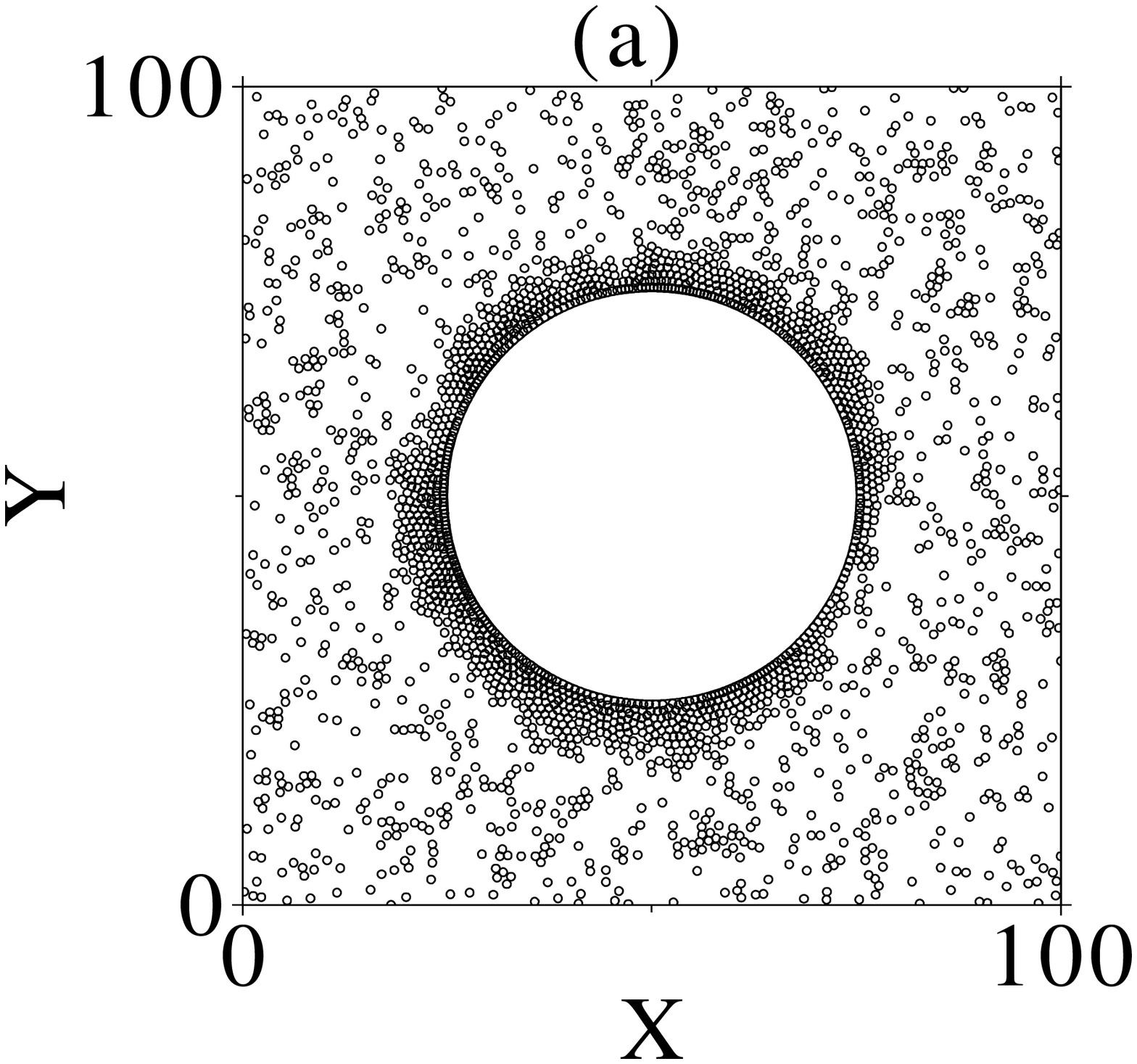} &
    \includegraphics[scale=0.2]{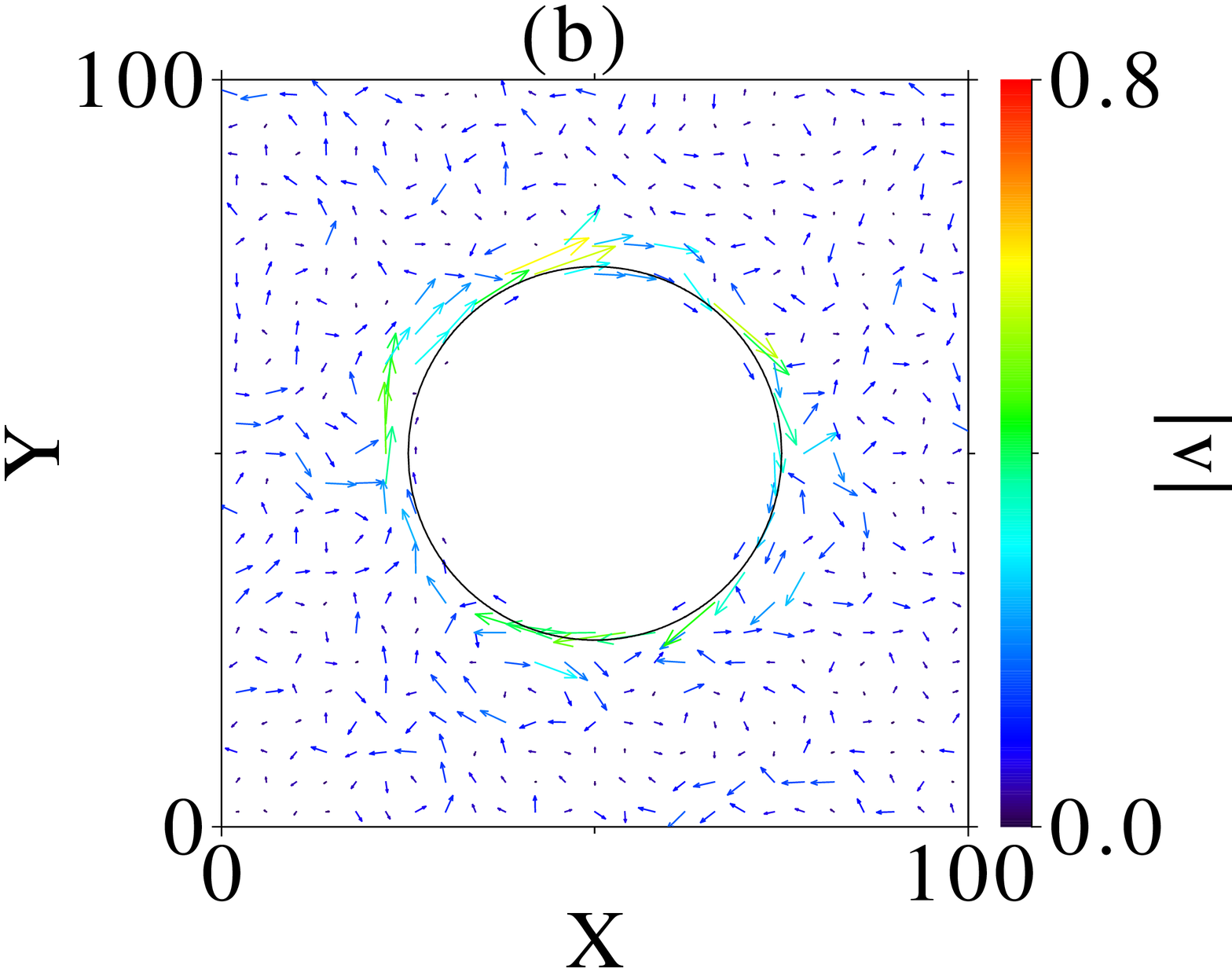}\\
    \includegraphics[scale=0.2]{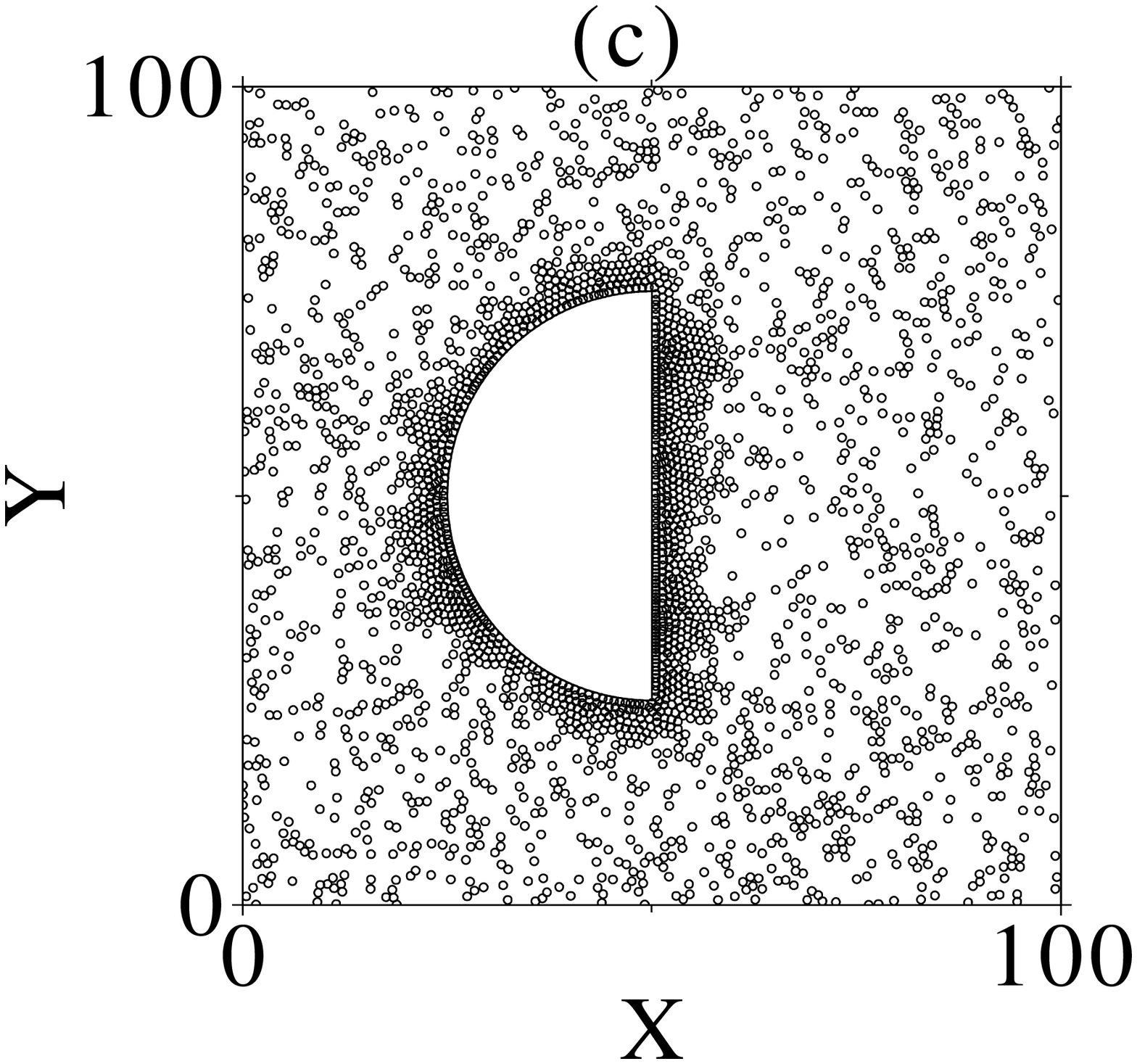} &
    \includegraphics[scale=0.2]{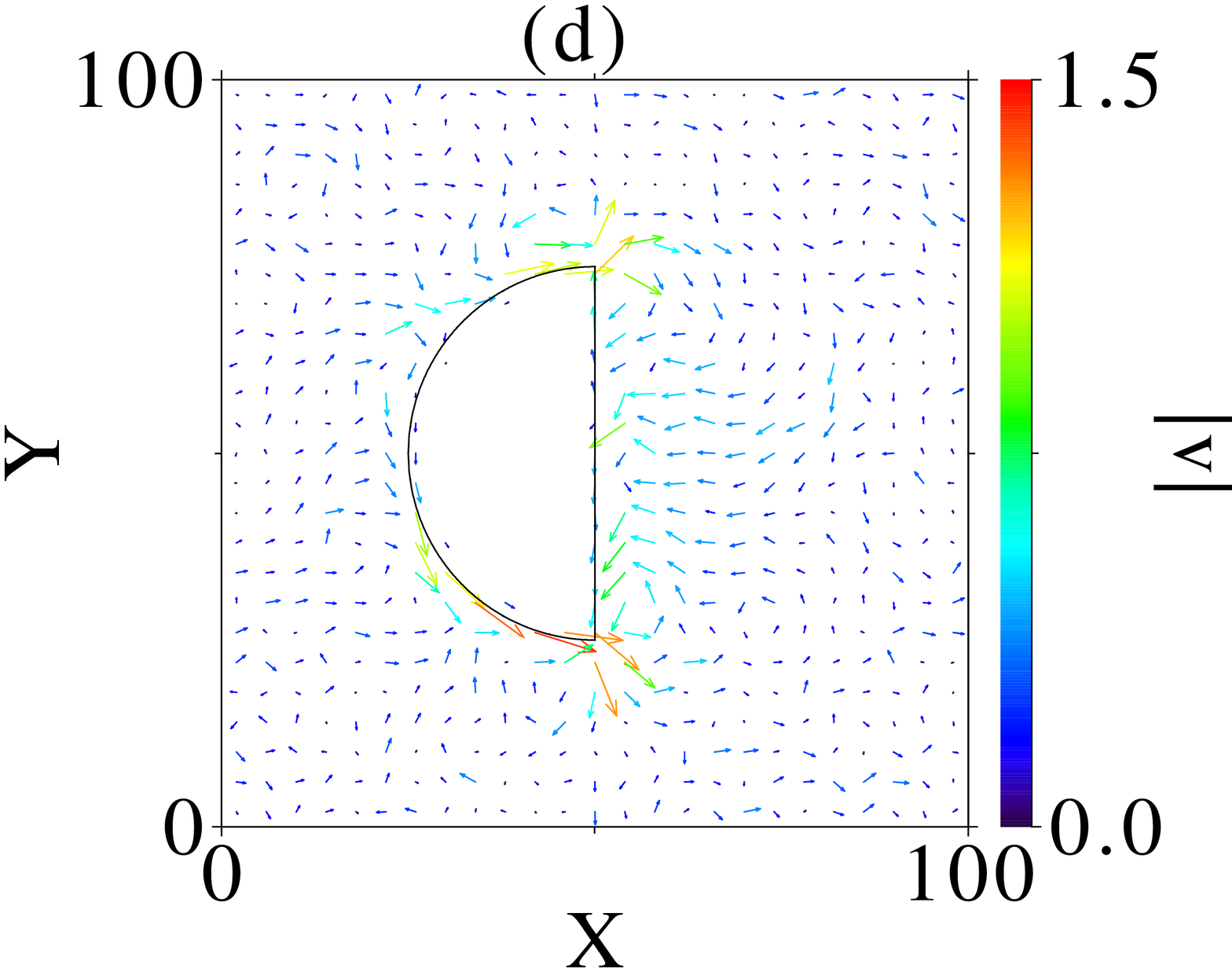}\\
  \end{tabular}
  \caption{(Color online) The configuration [(a) circle, (c) half-circle] and the average velocity field [(b) circle, (d) half-circle] of SPP in the presence of a single obstacle. Other system parameters are $D=50$, $\phi=0.596$, $\eta=0.005$.\label{fig1}}
\end{figure}

We start by discussing qualitatively how the presence of convex obstacles affects the dynamics of the swimmers. In Fig. \ref{fig1} we show the average velocity field and the corresponding snapshot configuration in systems with a single obstacle. Common features are that particles aggregate around the obstacles, and they follow a vortex-type motion around the curved surfaces with a direction (clockwise or counter-clockwise) chosen spontaneously. Notice that the swimmers strongly overlap near the obstacle boundaries, rendering very dense clusters there. This occurs due to the low value of the hardness of the linear spring interaction, $\kappa$. Larger $\kappa$ values yields weaker contact overlaps, since particles repel each other more strongly, although there would be still a large number of particles close to the obstacle. Care should be taken, though, when increasing the value of $\kappa$ regarding the value of the time step $\Delta t$: it should be decreased when $\kappa$ is increased, so that two swimmers do not overlap, even modestly, when they perform their inherent displacement, $v_0\Delta t$, since the repulsive force would be very large, forcing particles large distances apart in a single time step. The velocity fields in Figs. \ref{fig1}(b) and \ref{fig1}(d) were measured for distinct time intervals, the former being $100$ times longer than the latter. We observe that, although particles attach to the obstacles, the vortex motion is sustained only on the circular one. In the half-circle obstacle, the vortices quickly die out, since there is a very small probability for SPP to change their motion sharply to follow the flat side when emerging from the curved side. On top of that, particles already in the flat side provide an additional barrier that prevent the vortex to form. In our model, we do not consider hydrodynamic interactions (which is believed to be responsible for the motions observed in Ref. \onlinecite{takage13}), which could force the disks to attach to the solid surfaces. In fact, we observed that a single swimmer trajectory in a lattice of half-circles differs little qualitatively from a trajectory realized in a lattice of circles with the same initial conditions. We checked such results for several values of the noise intensity ($\eta=0.0001,0.01,1$) and UCL of the obstacle lattice ($a=10,100$). Since there can be no net transport in a lattice with circular obstacle (it is completely symmetric), a single swimmer would not yield a non-vanishing current in a lattice of half-circles. Therefore, these features appear only as a collective effect, where contacts among the swimmers are key to clustering \cite{fily12,palacci13}. Notice that aggregation is facilitated by the convex obstacles, since we always observed some degree of clustering around them.

\begin{figure}[h]
      \includegraphics[scale=1.3]{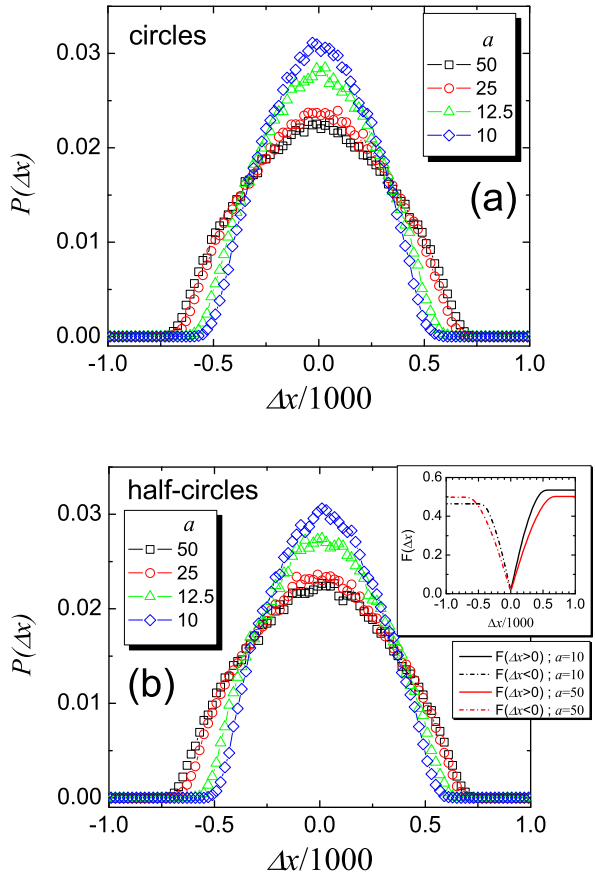}
  \caption{(Color online) Normalized particle displacement probabilities along the $x$-direction, $P(\Delta x)$ for (a) circular and (b) half-circular obstacles. In both cases the obstacle diameter is $D=5$ and $\phi=0.244$, $\eta=0.005$. The cumulative distributions for positive $F(\Delta x>0)$ (solid lines) and negative $F(\Delta x<0)$ (dot-dashed lines) displacements in the lattice of half-circle obstacles are shown as inset in (b) for $a=10$ (black) and $a=50$ (red, grey). \label{fig2}}
\end{figure}
In order to gain additional insight of the stationary state, we present in Fig. \ref{fig2} the normalized probability function for a particle to perform a horizontal displacement $P(\Delta x)$ in a lattice of obstacles as a function of $a$. The curves were obtained in the long time limit, enough for a particle to cover a distance equivalent to $|\Delta x|=1000$. 
In general, the curves are more narrowed and peaked around $\Delta x=0$ for smaller $a$, which means that more particles have their motion more restricted for denser lattices as a consequence of the accumulation of particles around the obstacles (aggregates). The probability for a particle to be caught in a given aggregate is proportional to the size of the aggregate. It is reasonable to assume that the aggregate size depends on the density of particles, obstacle size, and noise intensity, but not on $a$ (we also checked the dependence of the aggregate size on $\eta$ by measuring the density profile around the obstacles, and found that it decreases with increasing $\eta$). However, while decreasing $a$, and keeping all the other parameters constant, the aggregates approach each other, decreasing the free space for the SPP to move, and restricting their long time displacement.

From Fig. \ref{fig2} we also notice that $P(\Delta x)$ depends on the shape of the obstacles. For half-circles, \ref{fig2}(b), the curves are non-symmetric regarding $\Delta x=0$, implying a preferred direction in the motion of the SPP (rectification). Quantitatively, in the case of half-circle obstacles the cumulative distribution for positive displacements, $F(\Delta x>0)$, inset Fig. \ref{fig2}(b), black curves, is larger than the one for negative displacements, $F(\Delta x<0)$ red (grey) curves, which indicates that particles drift in the $+x$ direction, producing a steady particle current. The difference between these two cumulative distributions is a measure of the strength of the particle current, and it increases with decreasing $a$. The probability function for displacements along the $y$-direction $P(\Delta y)$ is essentially indistinguishable for both obstacle shapes due to the symmetry along the $y$-direction.
 
From the previous discussions we learn that: $i$) SPPs present a vortex-type motion around convex symmetric obstacles even in the absence of hydrodynamic effects. Such a motion is not observed for a single SPP, but it is a consequence of the aggregation of SPPs around the obstacles; $ii$) a steady particle current rises spontaneously in a lattice of non-symmetric convex obstacles, and in the absence of an external field. 

%
\subsection{Half-circular obstacle} 
\begin{figure}[h]
    \includegraphics[scale=1.4]{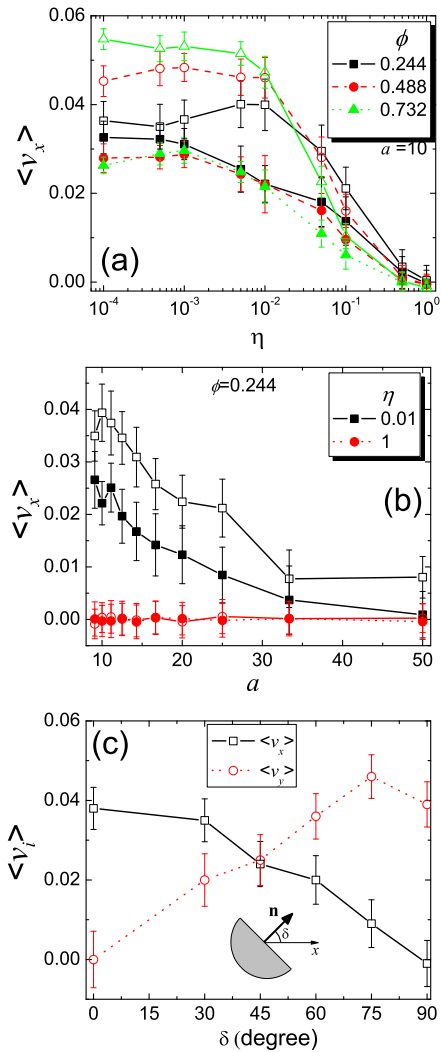}
  \caption{(Color online) (a) The average $x$-component of the drift velocity $\left<v_x\right>$ of the SPPs (in a lattice of half-circles obstacles) as a function of noise $\eta$ (log scale) for $a=10$, two obstacle sizes $D=5$ (solid symbols) and $7.5$ (open symbols), and three different area fractions $\phi$. (b) Same quantity, but now as a function of $a$, for two obstacle sizes $D=5$ (solid symbols) and $7.5$ (open symbols).(c) $x$- and $y$-component of the drift velocity as a function of the orientation angle $\delta$ of the half-circle obstacles with respect to the $x$-direction (inset).\label{fig3}}
\end{figure}

Now we take a closer look at the behavior of SPP in the array of half-circles by systematically studying the average drift velocities $\left<v_x\right>$ and $\left<v_y\right>$ in terms of $\eta$, $a$, $D$, and $\phi$. In general, our measurements yield only positive $\left<v_x\right>$, indicating that SPP move along the direction normal to the flat side of the obstacles ($+x$-direction), while perturbations along the $y$-direction are symmetric, resulting in $\left<v_y\right>=0$.  In Fig. \ref{fig3}(a) we present $\left<v_x\right>$ as a function of the noise $\eta$ for $a=10$, three distinct area fractions $\phi=0.244$ (black squares), $0.488$ (red circles), and $0.732$ (green triangles), and two different obstacle sizes, $D=5$ (solid symbols) and $D=7.5$ (open symbols). The ${\eta}$-dependence of $\left<v_x\right>$ is fairly insensitive to $\phi$. In general, low noise favors higher $\left<v_x\right>$ (which seems to reach a plateau for $\eta\leq0.001$), while $\left<v_x\right> \rightarrow 0$ for $\eta \gtrsim 1$. In addition, we observe that $\left<v_x\right>$ increases with increasing obstacle size $D$, but this is noticeable only for small noise ($\eta\lesssim 0.1$). For large noise values ($\eta\gtrsim 0.1$) $\left<v_x\right>$ presents little dependence on both $\phi$ (related to the density of SPPs)  and $D$. In Fig. \ref{fig3}(b), $\left<v_x\right>$ is presented as a function of $a$ for $\phi=0.244$, two distinct values of the noise $\eta=0.01$ (black squares), and $1$ (red circles), and obstacle size $D=5$ (solid symbols), and $7.5$ (open symbols). For large noise ($\eta \geq 1$), the drift velocity vanishes for any value of $a$. Such a behavior was also observed for the other values of $\phi$ considered in Figs. \ref{fig3}(a). For small noise ($\eta=0.01$), the drift velocity $\left<v_x\right>\neq0$, and it increases with decreasing $a$. We showed earlier that large noise disfavors aggregation, while small noise favors the accumulation of SPP around the obstacles. Since the results presented in Figs. \ref{fig3}(a),(b) relate higher $\left<v_x\right>$ with smaller $\eta$, and smaller $a$ or larger $D$, our findings indicate that cluster formation around the obstacles is essential for the occurrence of rectification (particle current) \cite{wan08,takage13}. The drift velocity increases with increasing $D$ (and decreasing $a$). Therefore, $\left<v_x\right>$  can be enhanced by bringing the obstacles closer (since decreasing $a$ is equivalent to increasing $D$). This trend is due to the overlapping of neighboring aggregates, which forces more particles in the +x direction. 

Up to here we considered the flat face of the half-circle obstacles perpendicular to the $x$-direction. In order to show the possibility to control the direction of the particle current, we present in Fig. \ref{fig3}(c) both $\left<v_x\right>$ (black squares) and $\left<v_y\right>$ (red circles) as functions of the angle $\delta$ between ${\bf{n}}$ and the $x$-direction [see inset in Fig. \ref{fig3}(c)]. The results clearly indicates the possibility to control the direction of the particle current by changing the orientation of the obstacles. Notice that $\left<v_x\right>$ and $\left<v_y\right>$ have, approximately, inverse variations with $\delta$, which indicates that the drift velocity is in a direction close to that of ${\bf{n}}$.



\subsection{Circular obstacles}
Our previous results and discussions indicate that no steady particle current (i. e., $\left<v_x\right>=0$) exists in a regular array of circular obstacles. Nevertheless, it is possible to induce rectification in the motion of SPP using this type of obstacle. Since medium asymmetry is essential for the motion rectification to occur, we can build a 2D square lattice of circular obstacles whose diameters are increasing functions of $x$ in order to produce such asymmetry. That this type asymmetry will lead to motion rectification can be seen as follows. From the results shown in Fig. \ref{fig1}, we may assume that the chance for a particle to get stuck in an aggregate is proportional to its size, and particles have a higher probability to move towards larger obstacles, in the present case, in the $+x$ direction. Therefore, the rectification we mean here is essentially particles moving from smaller towards larger obstacles. However, we observe that in the long time limit $\left<v_x\right> = 0$ in all cases studied (this ocurred even for periodic boundary conditions, since for solid ones this quantity would vanish). This can be explained by the fact that swimmers will move in a particular direction as long as there is a larger obstacle ahead in that direction, which can only occur in an infinite, non-periodic lattice. Hence, this rectification effect is a transient phenomenon in a finite lattice. Nevertheless, the steady state of this system, although it has a vanishing particle current, displays a non-uniform density profile along $x$-direction, because when particles reach the boundary of the system (either periodic or solid), in which there are no more larger obstacle for them to reach, they do not return to the smaller ones, precisely by the reason we pointed out above, they have a higher probability to get stuck in higher aggregates, in other words, around larger obstacles. 

From these considerations, we studied the density profiles as functions of the same quantities as before (noise intensity, $\eta$, unit cell length, $a$, and obstacle size, $D$, and periodic boundaries) and observed that more particles aggregate around larger obstacles (which can be interpreted as stronger particle separation, similarly to what was seen in \cite{drocco12-02}) for smaller noise intensity, and denser lattices. We also observed that, for large noise values, $\eta\geq1$, the density profile is inverted, and swimmers tend to move towards the smaller obstacles, while producing hardly any aggregation around any obstacle. We interpret this result on account of the need for the particles, when moving under such large noise values, to have a more free space to move (since they change their motion direction strongly in only a few moves), in comparison to the situation seen at low noise values, therefore accumulating, preferentially, in the region where there are the smallest obstacles (but not around the obstacles themselves, as seen in fig 1(a)).

\section{Conclusions}

We reported numerical results on the behavior of self-propelled particles in regular arrays of convex obstacles: either half-circle or circles. We showed that such an environment provides a means to rectify the swimmer motion. In the half-circle lattice, this rectification yields a finite drift velocity in the direction of the normal to the flat side of the obstacle, and homogenous spatial density. The drift velocity depends on density (area fraction), noise magnitude $\eta$ (low $\eta$ yields high speeds, and vice-versa), and density of the obstacles (larger rectification is found for denser lattices). We also showed that it is possible to control the direction of rectification (steady velocity) simply by adjusting the orientation of the half-circles obstacles. 
In the lattice of circular obstacles, rectification appeared when a size gradient in the diameter of the obstacle was considered. It is a transient effect which produces a migration of SPP towards larger obstacles, resulting, in the long-time limit, in a vanishing drift velocity, and a non-homogenous particle density. We found that the swimmers move towards the denser part of the lattice (larger obstacles) for low noise, while this tendency is reversed for high noise. We expect that our results open the way for new rectification (or separation) devices of active matter based on simple and symmetric lattices of convex obstacles. 


This work was supported by the brazilian science foundations CNPq, CAPES, FAPESPA and
FUNCAP (PRONEX grant).


\end{document}